\RequirePackage{fix-cm}
\documentclass[smallextended]{svjour3}       
\smartqed  

\usepackage[noadjust]{cite}
\usepackage{enumerate}

\usepackage{graphicx}
\usepackage{array}

\usepackage{tabulary}

\usepackage{caption}
\usepackage{subcaption}
\usepackage{color}
\usepackage{todonotes}

\newcommand{\av}{{\em AliCe-ViLlagE}}

\begin{document}
%
\title{The Collaborative Virtual Affinity Group Model:\\ Principles and Design}

\author{Ahmad Al-Jarrah \and 
		Enrico Pontelli \and
		Clinton Jeffery
	}
	
\institute{A. Al-Jarrah \at
			Al-Balqa Applied University\\
			Ajloun University College\\
			Applied Science Department\\
			\email{aljarrah@bau.edu.jo} 
			\and
			E. Pontelli \at
				New Mexico State University\\
				Department of Computer Science\\
				\email{epontell@nmsu.edu} 
				\and
			C. Jeffery \at
			University of Idaho\\
			Department of Computer Science\\
			\email{jeffery@uidaho.edu}}

\maketitle

\begin{abstract}
The problem addressed in this paper is the challenge arising in enabling collaborative
learning in the context distance education models. While research has made quantum leaps
in the development of both effective collaborative pedagogical models as well as online 
learning environments, the research at the intersection of these two areas has been scarce.
This paper presents the design for a new collaborative virtual model, named 
\emph{Collaborative Virtual Affinity Group model (CVAG),} which is an extension of 
the successful Affinity Group Research (ARG) model.  
 The new model provides an integration of the principles underlying ARG with the traditional
 principles of  virtual learning environments. 
The CVAG model is explored in the context of introductory Computer Science courses---where students
are focused on learning the basic principles of computer programming.  
\end{abstract}

\section{Introduction}

\subsection{Motivation}
The role of collaborative learning \cite{col0,col1,col2} has been widely explored in the
pedagogical literature and practices. 
The theoretical foundations for collaborative models of learning, especially
in scientific areas like mathematics and computer science,
can be found in the seminal work on  
\emph{constructivism} \cite{davis90,ernest95}, where the learning process is
viewed as active participation of the student, engaged  in a process 
of recursive construction of knowledge. The student builds
 new cognitive structure from the combination of his/her existing knowledge,  the  experience
imparted by the instructor, and the interaction with peers. In particular, these considerations span from social constructivism in teaching and learning \cite{vygotsky}, bringing into account the role played by the immersion of the student in a community of learning;  the physical and social experience of interacting with other students  provides a valid scaffold to the learning
experience.

The role of collaboration in the learning experience has been shown not only to 
provide an effective pedagogical instrument, but also to facilitate the engagement
of students from traditionally underrepresented backgrounds \cite{div1,div2}. The role 
of collaboration is well aligned, for example, with traditional values found in 
Hispanic communities. Hispanic
families are characterized as valuing  family needs over individual needs \cite{div3}, strong loyalty and reciprocity
among family members, and  high regard for behaving respectfully and in ways that promote
harmony \cite{div4,div5}. Hispanic students share 
 a collectivistic culture, which values the role of 
 shared responsibilities, and where accountability is towards the
 group \cite{div6}.
 
Computational tools and methods have gained a central role in supporting collaborative learning; innovative 
computational instruments have been proposed to promote interaction between students, provide challenges, and
facilitate data collection and analysis for the benefit of instructors \cite{dist0,dist1,dist2,dist3}. Indeed,
an entire research field has developed focused on computer-supported collaborative 
learning (e.g., \cite{dist4,ndist,ndist1,ndist2}).

The role of collaborative learning has gained momentum in a variety of scientific disciplines. In the
area of Computer Science, collaborative learning models are at the foundation of several
successful pedagogical methodologies (e.g., peer-led team learning \cite{pltl}, pair-programming \cite{pair}). 
The \emph{Affinity Research Group (ARG)} model \cite{ARG} is a formal model developed, in the
context of Computer Science education, to promote
structured research groups composed of undergraduate students, where students with diverse backgrounds and levels of preparation are
made active participants in both the research and learning experience. ARG has been shown to be
a very effective group model in promoting learning, participation, engagement and inclusion \cite{arg1,ARG2008}.

\subsection{Summary of Contributions}

The success of the ARG model has been validated in a variety of settings, within different research
areas and with students from different backgrounds and levels. Nevertheless, the ARG model has been
applied exclusively in situations that are {\bf spatially} and {\bf temporally} bounded. The \emph{spatial}
bounds lead to applications of ARG that are entirely face-to-face---i.e., a team of students working 
on a project in class. The \emph{temporal} bounds 
imply that the operations of an ARG team are synchronous, and often limited to a specific time
period (e.g., a lab session). Both bounds represent 
obstacles to a broader use of ARG, to support research and learning teams that 
are distributed. Such transition, in particular, is critical to enable the use of ARG in the context
of distance education and online courses. 

In this paper, we propose the design of the \emph{Collaborative Virtual Affinity Group (CVAG)} model. CVAG represents
an extension of ARG that relies on the use of \emph{Collaborative Virtual Environments} to promote the formation
and operation of distributed affinity groups. In order to guarantee feasibility, we explore the design and development of 
CVAG in the focused context of affinity groups of students learning introductory computer programming---or engaged
in research projects associated with basic programming. 
Hence, CVAG can be defined as a collaborative virtual model that supports 
distributed affinity groups of students to learn computer programming in a collaborative manner.

\section{Brief Literature Review}
\label{BLR}

\subsection{The Affinity Research Group (ARG) Model}

	The Affinity Research Group model (ARG) is a model to structure and operate
	 undergraduate research teams~\cite{ARG2008}. Specifically, 
	it is a comprehensive model for creating and maintaining dynamic, productive, and inclusive research groups.
	 ARG creates a cooperative environment designed expressly to allow each member (e.g., 
	faculty mentors, undergraduate students, and graduate students) to flourish~\cite{CUR}.  
	In ARG, research goals are shared among group members, in addition to academic and professional development goals. This affinity is the basis for promoting cooperation in achieving goals. The contributions of each group member are essential to the group's success---which is defined as reaching a set of goals which have been agreed upon by the entire group. 

ARG was developed by researchers at the University of Texas at El Paso, under funding 
from the National Science Foundation~\cite{ARG}. The motivation behind the name ARG can be found in 
 the fact that  research group members that use the ARG model  develop an {\em affinity} for the particular research topic~\cite{ARG}. 
 The creators of ARG found that undergraduate research is very important for helping students to identify the skills and knowledge they need to successfully participate in research projects, and appreciate the rewards of graduate study. ARG has been formally
 studied and  shown to increase the ability to do research, improve 
 students' self-efficacy, and enhance skills in presenting and defending  ideas~\cite{CUR}. There is empirical and experimental evidence showing that the use of  ARG leads to 
 improved retention and graduation of students in Computer Science, Electrical and Computer Engineering, especially students
 with deficiencies and  low levels of confidence  \cite{ARG2008}.

	The philosophy behind the ARG model is to maximize each student's ability to reach his/her potential. It is built on the principle of consciously developing and training researchers in a cooperative environment, offering to the team members specific roles and rules of operations. The dynamic assignment of roles to team members is aimed not only 
	to enable the team's progress towards its goals,
	but also as an  intentional development of the skills and sense of responsibility of each team's member. 
	Throughout the lifetime of the research group, the members alternate in the different roles, including the 
	roles of leadership. The dynamic reassignment of roles allows each team member to develop different skills and
	ensures that each member has the opportunity to contribute to the team goals. It also ensures that each member is
	accountable for her/his contributions to the team activities. 
	In particular, in the context
	of educational applications of ARG, the leadership roles are shared between students and faculty members.
	The ARG model  began as a mechanism to enhance retention of students from traditionally underrepresented groups, such as female students, in computing disciplines. The model has evolved since, making it  suitable for a variety of undergraduate groups~\cite{ARG}. 
 
  The ARG model strives to structure group interactions in a way that fosters positive interdependence and promotive interaction, generally yielding the following outcomes:
  \begin{itemize}
  	\item Establishment and maintenance of cooperative groups and subgroups;
  	\item Achievement of research deliverables; and
  	\item Self-development of group members, who acquire teaming and research skills and progress in cognitive and emotional development.
  \end{itemize}
  Using structured tasks and activities, ARG encourages students to: 
  \begin{itemize}
  	\item Develop domain expertise,
  	\item Understand and appreciate the research process and its practice, and
  	\item Acquire team, communication, problem-solving and high-level thinking skills that will make them effective leaders and successful in research, academia, and industry.
  \end{itemize}
  Let us contrast ARG  with more traditional models to structure and operate research groups. The traditional models tend to rely
on hierarchical/pyramid structures,  where the layers represent a decreasing order of expertise and authority.
While ARG does incorporate aspects of the traditional models (e.g., peer-to-peer relationships between
team members), it also differs deeply in several  important aspects, as summarized in Table \ref{tab1}.

\newcolumntype{L}{>{\centering\arraybackslash}m{4cm}}

{ \renewcommand{\arraystretch}{2.5}
\begin{table*}[htbp]
{\footnotesize
\begin{center}
\begin{tabular}{|l|L|L|}
\hline\hline
{\bf Category} & {\bf Traditional Model} & {\bf ARG}\\
\hline\hline
\emph{Leadership} & 
The leadership position is reserved to the faculty mentor and s/he is only the person who provides direction to the other group members
& 
Each student acquires the skills to be a leader and expert in some aspect of the research that the group is doing as a whole\\
\hline
\emph{Tasks} & 
	 Members are often concerned about the progress of their individual project &
		Members are concerned about the progress of the team's project\\
	\hline
\emph{Participation} &
		Only the best and brightest are recruited, and undergraduate students are rarely included
	 &
		Heterogeneous membership is encouraged, and undergraduate students are recruited\\
\hline
\emph{Skills} & 
		``Necessary'' research and technical skills are taught&
		Research, technical, team and professional skills are emphasized and explicitly taught\\
	\hline
\emph{Development} &
			Professional skills are assumed &
			Professional skills are developed through structured activities\\
		\hline
\emph{Environment} &
		Environment is controlled by the research leader and may be competitive& 
			Cooperative environment is a key part of the model and is encouraged and developed
		\\
	\hline
\emph{Improvement} & 
		Process improvement is not practiced or is as hoc &
		Process improvement is part of the model \\
	\hline\hline
\end{tabular}
\end{center}}
\caption{ARG vs. Traditional Group Models}
\label{tab1}
\end{table*}}

Furthermore, ARG can be applied to virtually  any type of groups, allowing for a broad
range of diversity between the group members (e.g., capabilities, interests, 
skills, education, culture, backgrounds, and experiences). This is thanks to  ARG's infrastructure, 
which  provides students with active mentoring and continuous
training in research.

\subsection{Collaborative Virtual Environments (CVE)}

There is an extensive literature exploring the nature and role of
\emph{Collaborative Virtual Environments (CVEs)}---viewed as multi user, distributed worlds~\cite{liebregt2005}.
The authors of~\cite{greenhalgh1997large} define CVEs
 as ``computer based systems which actively support human collaboration and communication within a computer based context". The author of~\cite{redfern2002} define CVEs as ``computer-enabled, distributed virtual spaces or places, in which people can meet and interact with others, with agents and with virtual objects".  The other definition of CVEs by~\cite{liebregt2005} is ``connected computer systems aimed at the fulfilling of a certain collaborative task within a generated 3-D virtual environment". 

A collaborative virtual environment provides a framework to enable collaborations among  a number of users, working on 
a common task in a spatially distributed scenario. According to \cite{greenhalgh1997large}, there are a number
of basic properties that underlie the design of most collaborative virtual environments:
\begin{itemize}
\item CVEs are multi-user computer-based systems which support geographically dispersed users;
\item CVEs provide the user the ability to collaborate and communicate in a number of different ways;
\item CVEs are virtual environments---i.e., they are based on a virtual space or world, where all activities are 
	performed;
\item Each user is explicitly represented within the virtual environment and visible  to other users by means of embodiment;
\item Each user is independent from the others and has the ability to make any movement within the virtual environment independently.
\end{itemize}
CVEs have found extensive use in the domain of online education, and a number of educational 
CVEs have been developed and discussed in the literature.  Liebregt~\cite{liebregt2005} studies six 
different CVEs to answer the following question: what is the potential of  CVEs  as learning tools? 
Table~\ref{tab:EX-CVE1}  contains a comparison between these six environments with their features and other characteristics. The study concluded that CVEs provide a number of
advantages in an educational setting: 
\begin{itemize}
\item They support the social awareness of university students;
\item They increase communication and discussion possibility on a wide scale;
\item They support constructivist learning on different subjects aimed at different age groups;
\item They increase information available to users;
\item They enable collaborations that culminate in sharing and creation of new knowledge;
\item They provide  virtual experiences that facilitate the  learning of challenging concepts using different 
learning strategies.
\end{itemize}

\begin{table}[htbp]
\centering
{\footnotesize
\begin{tabular}{||>{\centering\arraybackslash}m{1.2cm}|p{3.7cm}|p{2.5cm}|p{2.3cm}|p{2.2cm}|p{1.7cm}||}
\hline \hline
\bf{Name} & \bf{Development reason} & \bf{Features} & \bf{Audience} & \bf{How to access} & \bf{Hardware requirements} \\ 
\hline \hline
CVE-VM & 
It is a part of the virtual museum projects & 
Text chat & 
Children and teen-age students &
Internet using a computer and browser &
No \\ 
\hline 
DeskTOP & 
Supports and promotes  collaborative learning in universities & 
Slider tool, audio and video, newsgroup tool& 
University students &
Internet &
No \\ 
\hline
DigitalEE and DigitalEE II & Uses a CVE in environmental education & 
3D virtual representation; text chat, voice communication, avatar &
Learners, virtual tourists as well as experts on the environment &
Desktop computer with Internet and mobile computer equipped with GPS &
Mobile computer equipped with GPS, digital camera \\ 
\hline
NICE &
Investigate how effective a CVE can be when used for learning and evaluation & 
Gestures (avatars), spoken word, text chat &
Children between 6 and 10 years old. & 
Desktop computer connected to the Internet & 
Digital camera, Special glasses, sensors etc. \\ 
\hline 
Round Earth & 
Builds according to NICE. 
Supporting the education on a difficult learning problem. &
As NICE &
As NICE &
As NICE &
As NICE \\ 
\hline 
Viras & 
How different factors of CVE impact on the social awareness. How students experience CVEs meant for increasing social awareness. &
Avatars using chat, sending messages and making gestures &
Students &
Desktop computer connected to the Internet. &
No \\ 
\hline \hline
\end{tabular} }
\caption{A comparison between six different environments}
\label{tab:EX-CVE1}
\end{table}

\section{The Theory Behind the Model}
In the last decade, a large number of  virtual and 
interactive technologies have been proposed to enable the learning of introductory programming; these 
technologies are designed to support a diversity of learning styles and backgrounds. At the same time, 
we have witnessed the extensive development of learning management systems that make it possible to support ``learning at a distance,'' meeting the needs of online and distance education.

In this paper, we present the design  of new collaborative learning model, referred to as 
the \emph{Collaborative Virtual Affinity Group (CVAG)} model. CVAG is designed to be used to teach students
introductory programming using a  virtual environment. The core ideas for the model are analogous
 to those of the  ARG model, but extended to enable their use in a distributed setting
using a collaborative virtual environment. In this section, we present first a general overview of the model,
followed by a discussion of  the main components of CVAG, as they derive from ARG, and how the components of ARG
have been adapted to enable their use in a distributed virtual environment. On the other hand, CVAG includes a number of collaborative environments components and properties. The final subsection discusses the CVAG as a collaborative virtual environment and how the CVE components mapped to be the essential components of CVAG.

\subsection{Overview}
Following the considerations made above, we can intuitively define CVAG as 
\emph{a collaborative virtual model that support affinity groups of students to learn computer programming in a collaborative manner.} The new model supports the process of collaborative learning, by inheriting from ARG the  features and components that support the collaborative learning process, improve learning experiences, and enhance programming skills. These features are combined with the benefits of collaborative virtual environments, such as  the ability for the group to work together and achieve  common goals, regardless of spatial location. 

CVAG shares ARG's emphasis on the development of structured working groups as a mechanism to create learning opportunities for the students, where students learn to use and integrate their knowledge and skills as  required for their research and work.  The design of the CVAG model illustrates the foundations of building an educational process to learn programming in a social context; CVAG provides a framework for each member of a team to learn, participate and share experiences and skills with members of the group. Each member is expected to interact and participate along the entire life-cycle of the learning experience---as
ARG emphasizes accountability and awareness that each team member's contributions are essential for the success of the whole
team;  any failure affects negatively not only the individual student, but  the whole group. In order to achieve such behavior in a truly distributed setting, where students may be located at different sites (e.g., at home, at different institutions), CVAG reinforces the  creation of customized collaborative environments to learn programming and to support collaboration, communication and social interaction, involving both students and instructors, and relying on shared virtual workspaces.

\subsection{Model Components: Elements of Collaboration}
CVAG extends ARG's components and merges them with the five elements of 
distributed collaboration: positive interdependence, face-to-face promotive interaction, individual and group accountability, professional skills and group processing~\cite{felder2007,ARG}. The five elements of collaboration are used to organize the work of the group, determine the methods that are used to ensure achievement of the group's goals, and decide what are the required features to facilitate the work of the group members. The success of the group depends on the success of each member in the group, therefore each member has a responsibility to complete their assigned task to support the success of the whole group. In CVAG, the five elements are enhanced to support the group work within a virtual environment. Let us elaborate on each of these collaboration principles to illustrate how they are captured in CVAG.

\subsubsection{Positive Interdependence}
Each member in the group understands that his/her contributions are essential for the success of the whole group and s/he has to finish his/her part successfully. In CVAG, following the ARG model, we rely on the 
use of \emph{roles} to scope the duties of each individual team member; roles are assigned at the beginning and dynamically revised throughout the execution of a project (e.g., a classroom project). Each role is supported by a number of features, which support the role functionalities (e.g., a note-taker receives  a special white-board to write notes about the work). 
CVAG gathers information of the work's progress and shows the amount of contribution made by each member. 
Such statistics assist the supervisor (e.g., an instructor, a team leader) to evaluate each member's contributions, to 
assess
the overall progress towards the goal (e.g., the completion of the homework), and make decisions on how to organize
the work process (e.g., introduce new lecture modules to fill recognized learning gaps). CVAG ensure positive interdependence by:
\begin{enumerate}
	\item Providing a shared workspace that all members can access, with guarantees of consistency, to collaborate on a shared   group project (in our case, the development of programming project in an introductory programming course);
	\item Each programming effort is divided into tasks, and each member has to finish a specific task to lead to the
		completion of the effort.
	\end{enumerate}

\subsubsection{Face-to-Face Promotive Interaction}

CVAG provides a learning environment for the group,  where members are enabled to share knowledge, help and encourage each other's effort to learn. In CVAG, a number of communication channels are
provided to the team members---i.e., text chat, audio, and video---and three types of meetings are supported
to allow students' interactions---orientation, small group meetings and large group meetings. We will describe these phases
in detail in a following section of this document. Intuitively, in a small group meeting, students have a conversation all the time while they work, share knowledge and help each other. In large group meeting, they share what they achieved and discuss the whole group progress and encourage each other to finish, and celebrate with members that finish their jobs successfully.

\subsubsection{Individual and Group Accountability} Achieving a team's common goal requires contributions from each member in the group. CVAG assesses the performance of each individual, by collecting information about each member's contributions for each type of transaction. The statistics can be  reviewed by selected members in the group and by the supervisor (e.g., the teacher).

\subsubsection{Professional Skills} In CVAG, students learn problem solving, while practicing how to 
operate effectively within  a team. CVAG's assignment of roles, with associated features and restrictions, is designed to scaffold the development of professional skills, such as time management, communication, note taking, summarization, accountability, and critique.

\subsubsection{Group Processing} Group members should decide how the group goals can be achieved and what is the best method to work together. At the beginning, a team leader or supervisor (e.g., a teacher) assigns  roles to each member in the group; nevertheless, these roles are dynamic. Therefore, group members and the supervisor can change roles, as needed to achieve the project goals and meet the expected learning objectives---e.g., ensure that each student actively expresses ideas and opinions. 
This type of change can happen during the small or large group meetings, and after discussing the results and work progress. Moreover, a meeting can be held to discuss and determine what are the required changes that can help to improve the  progress. 

\subsection{Model Components: Principal ARG Components}
\label{ARG_maped_comp}
The other components of CVAG are the result of mapping the essential elements of ARG into CVAG: core purpose, students connectedness and management scheme. These three main components are enhanced and 
modified to be suitable to be used in CVAG.

\subsubsection{Core Purpose} In ARG, one of the main  initial steps is to identify the core purpose and core values of the group. This step is mapped directly to CVAG. Core purpose determines the reason for the  group's existence. During  the orientation
phase, the supervisor defines the core purpose for the group; the core purpose provides to the team a  clear vision to drive the group's planning, in addition to offering metrics to assess progress. The group will use the core purpose as a guidance and inspiration to work together to achieve the common goals. Moreover, the core purpose is a good start for the group to define the core values. CVAG embraces the same three core values as  ARG: student success, cooperation, and excellence~\cite{ARG}. 

\subsubsection{Student Connectedness} Studying programming while working with a group is not a familiar technique for most new students. Therefore, the supervisor has to provide a clear description of each individual task and the connection of how the different tasks contribute to the overall project goals.  The description should be available for all group members and everyone can access it any time. CVAG provides a space for the supervisor to publish the clear description. In the orientation, the supervisor explores students' perspectives and concerns, and establishes the initial tasks based on the students' experience and skills. Students are given the opportunity to discuss the requirements within the virtual environment.

\begin{figure}[htbp]
	\centering
	\includegraphics[width=.85\textwidth]{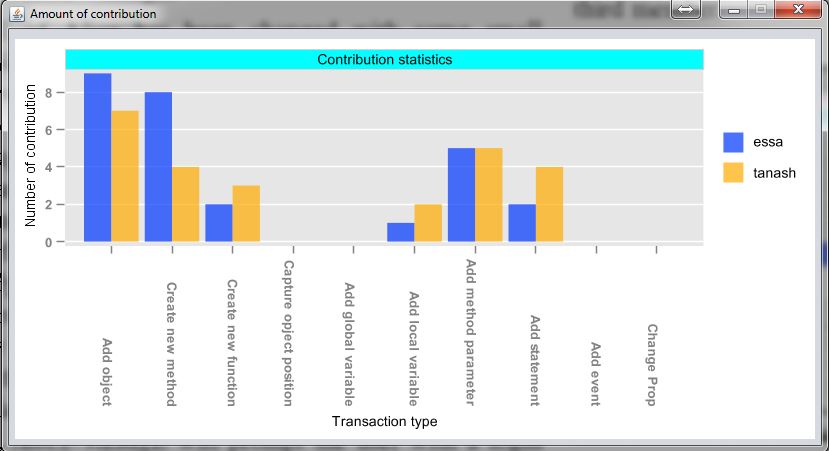}
	\caption{Example of the type of statistics that CVAG can provide to the supervisor and students---the example shows the contributions of two students}
	\label{fig_stat}
\end{figure}

\subsubsection{Management Scheme} The concepts of \emph{team} and \emph{team management}  are critical components for the success of the CVAG model. Learning to program and applying programming skills to specific problems are challenging tasks. They require students to learn and practice a number of skills in order to produce a good quality software artifact. In addition, learning to program within a team requires additional skills, in terms of 
communication, accountability and interaction patterns. The supervisor {\it defines the dependencies and time-line} for each part in the project. CVAG assists  the supervisor in determining the work steps, dividing the programming tasks into parts,
distributed as stages along a time-line. In particular, the CVAG supports the work for the group to be mentored: 

\begin{enumerate}
\item The supervisor determines the stages and the time-line for each stage.
\item Roles are assigned to students  at the beginning (e.g., the roles can assigned randomly, by vote, based on
	students' preferences, etc.). Roles are dynamic and they  can be changed at any time. 
\item The system collects information about the  progress in the project,  by saving data about each transaction
	performed (see Figure~\ref{fig_stat}) (e.g., type, time, who made it, etc.). 
\item The collected data can be used by the supervisor, students, and the system. The supervisor can review the  progress and evaluate the students' contributions, modify roles and determine learning progression and needed learning
interventions. Students can use the  information to encourage each other, discuss and resolve common challenges, and identify
alternative roles that might be considered.
\end{enumerate}

\begin{figure}[htbp]
\centering
\includegraphics[width=3in]{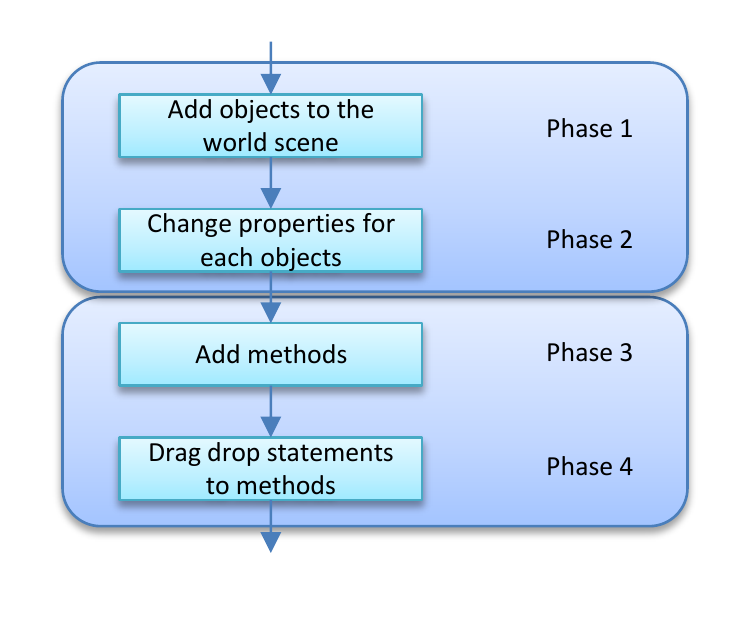}
\caption{An example for a number of phases to build a world in Alice programming language}
\label{fig_del}
\end{figure}

The other important thing that the supervisor has to clearly define are  the {\it expected deliverables}. Each task is divided into a number of subtasks, typically leading to a hierarchical decomposition of the overall task. Each level in the hierarchy is assigned to a specific working phase. Therefore, students cannot proceed to next phase (subtask(s)) until they have completed the  current phase or some relevant part of it. The example in figure~\ref{fig_del} shows that in phase 1, students need to add objects to the world scene, then change the properties for each object in phase 2. These two phases can be overlapped, where students can justify the properties for an added object before it is complete adding other objects (the same thing for phase 3 and 4). Whatever technique is used (finish adding objects first then change properties or changing properties while adding objects), students can deliver a world scene that contains the required objects after completing the first phase, and after phase 2, they can deliver how the world will be after they finish justifying properties for all objects, etc.

In CVAG, all the students will operate on  the same workspace (i.e., viewing the same version of the program being
developed); this allows the supervisor to maintain a view of the overall work been performed in each group. The CVAG 
provides also alerts to the group when the time has come to make a submission. The group leader can submit the deliverables when the team has completed the work or when a deadline has been reached. At the time of submission, the  CVAG will clone a version of the workspace for the supervisor and alert him/her that the  group has performed  a submission. .

{\it Activities} and {\it meetings} are two other important components of CVAG. CVAG requires the activities and meetings to be structured in such a way to make them suitable to be used in a virtual environment. In the next two sections, we will discuss activities and meetings in detail.

\subsection{Group Activities and Members' Roles}

CVAG is an environment to teach students the foundations of  computer programming; as such, it is important for the model
to capture the fundamental activities associated to the process of software development, especially in the perspective of exposing
inexperienced students to such processes. Thus, the expectation of CVAG is not only to serve as an instrument to support collaborative learning
of introductory programming, but also as a tool to introduce students to the formal and informal aspects of team-based software development.

CVAG supports four types of activities that students can practice: analysis, design, coding, and testing. Let us review these four tasks and the
relation between the activities and the available \emph{roles} provided by CVAG.

\subsubsection{Analysis}
{\bf Analysis} focuses on the understanding of the initial activities required of the students. After students receive their tasks, each group starts  by analyzing the task requirements. The \emph{analyzer} is a role that is initially assigned to all of the students; the only distinction is the
presence of a \emph{lead analyzer}, which is assigned to one member, who will serve as  the leader for the group. 

\begin{example}
In \av~\cite{7044089}, a collaborative virtual environment for pair programming in Alice \cite{alice}, at the end of the analysis activity, it is expected that the  students will be able to provide a comprehensive design of how to create a desired world scene, along with a breakdown of the design in 
terms of a list of required Alice objects, a list of properties with their values for each object, and a list of methods. A short description is also expected to be added for each item in the list, to explain how each element contributes to the overall design. Students can communicate  using any communication media available, to discuss the task, ask questions to each other, and make a decision about the overall design and each list. 
\end{example}

CVAG, should provide features to support the activities of all students in this phase. The leader student has the ability to set up the lists and other work areas, where the other students in the team can interact and exchange ideas. When the leader adds any item to a list, it is shared directly with everyone in the group. CVAG's features for the  analyzers include support for team discussion, the ability to take personal and shared notes and collaborate in the development of the lists of objects, methods and attributes that are established by the lead analyzer.

\subsubsection{Design}
The second activity is the {\bf design} activity. To make our discussion more concrete, let us focus the role of CVAG to the problem of designing applications for introductory programming environments, where students construct stories, scenarios, and virtual worlds. Students in this phase work to create simple program scenarios. Therefore, at the end of these activities, the  students are expected to submit a script that contains the use cases. The script usually contains answers for a number of questions, such as:
\begin{itemize}
\item How users will use the application?
\item What are the different functionalities provided by the program?
\item What are the series of events in the  story?
\end{itemize}

\begin{example}
Let us assume that  students are asked to tell a story using \av. The students will have to develop
a script for the story. All of the students in the group work together to create the script---thus, they all act 
as \emph{designers.} Everyone in the group will write his/her notes about the design and contribute to the discussion
leading to the script design. The team  leader (i.e., the \emph{lead designer}) will summarize the final decisions concerning the script.
The features provided by CVAG allow the leader to incrementally write the script and  share automatically with 
the whole group. The other members in the group can view the script but they cannot edit it. They can comment and participate in the discussion (using text, audio and/or video). 
\end{example}
In both the analysis and design activities, the leader role can be 
reassigned at any time according to the supervisor and the group's decisions.

\subsubsection{Coding and Testing}
{\bf Coding} and {\bf testing} are the two most time consuming activities for students learning to program. The work on these activities can be done in one phase or can be divided into two or more sub-phases according to the supervisor's requirements and how s/he defines the deliverables. 

\begin{example}
The supervisor asks students to create an
interactive  story. The supervisor can divide the work in two phases: 
{\bf (1)} the story without  user's interactions and
{\bf (2)} The story after adding  users' interactions. 
The main roles in these activities are {\it author (driver)} and {\it navigator}. While the driver is in charge of
writing the actual code, the navigator serves as reviewer, debugger and advisor. In CVAG, the supervisor is 
allowed to establish an arbitrary number of  drivers and navigators in the team.  The author is the member that works on creating the application by adding the required objects, changing properties, adding methods, etc.; the navigator follows the author. Testing can be performed at any time. When any member has any notes about a test, s/he can transfer the information directly to the group using any communication media (text chatting, audio, or video). 
\end{example}
Observe that the coding and testing activities 
can be alternated in an arbitrary manner, as decided by the team members.

\subsection{Analyzing Member Roles}
Leader, analyzer, designer, author and navigator are the  roles that we mentioned previously in relation to specific
activities. Table~\ref{tab:Roles} contains a list of roles that have been described in  ARG, along with the identification of which roles are also
present in CVAG. A brief description of each role is provided in Table \ref{tab:role_def}.  In CVAG, students collaborate synchronously to complete the required task. This enables team members to dynamically switch between different roles. For example, the concurrent development of code suggests the presence of multiple authors; similarly, an author can hold, at the same time,
the role of  explainer. Similarly, the group leader can act as an author at the same time as s/he is the leader for the group, therefore,  CVAG allows him/her to take notes, send messages, etc. as a leader, and allows him/her to add, modify, edit the code, etc. as an author. 
CVAG allows students to choose what  roles they act at any point in time---or such roles can be predefined and pre-assigned
 by the supervisor (e.g., the teacher). 

\begin{table}[htbp]
	\centering
	\begin{tabular}{||l|c|c||}
		\hline\hline
		{\bf Role } & {\bf ARG} & {\bf CVAG} \\
		\hline
		Leader & No & Yes \\
		\hline 
		Analyzer & No & Yes \\
		\hline
		Designer & No & Yes \\
		\hline
		Author & No & Yes \\
		\hline
		Time keeper & No & Yes \\
		\hline
		Explainer & Yes & Yes \\
		\hline
		Recorder & Yes & No \\
		\hline
		Gate-keeper & Yes & No \\
		\hline
		Direction giver & Yes & No \\
		\hline
		Clarifier/paraphraser & Yes & Yes(*) \\
		\hline 
		Summarizer & Yes & No \\
		\hline
		Accuracy coach & Yes & Yes (**) \\
		\hline
		Understanding checker & Yes & No \\
		\hline
		Elaborator & Yes & No \\
		\hline 
		Perspective-taking roles & Yes & No \\
		\hline
		Idea criticizer & Yes & No \\
		\hline 
		Justification asker & Yes & No \\
		\hline
		Differentiator & Yes & No \\
		\hline
		Integrator & Yes & Yes \\
		\hline
		Extender & Yes & No \\
		\hline
		Prober & Yes & No \\
		\hline\hline
	\end{tabular}
	\begin{flushleft}
		(*) This role can be useful to be used in case of students need more clarification about the project.\\
		(**) This role known in CVAG as Navigator. 
	\end{flushleft}
	\caption{The transferred roles from ARG to CVAG in addition to the new added roles to CVAG}
	\label{tab:Roles}
\end{table}

CVAG's ability to allow dynamic modification/reassignment of roles creates opportunities for very flexible
programming group models.  There is a significant value in allowing this type of extension:
\begin{enumerate}
\item It allows the team members to alternate in the roles that are provided (e.g., upon switching phase in the project)---this has a strong pedagogical value, as it allows students to experience different duties and activities; this also contributes to the development of a greater understanding of the overall team work and greater level of accountability. It is expected, for example, for a teacher to force the rotation of roles to ensure that each team member effectively contributes to the 
overall project. 
\item It allows more than one member to concurrently act in the same role, thus allowing students to share and exchange
experiences and act as peer-mentors. 
\end{enumerate}
Each  students enters a project as an analyzer and, after the analysis phase is completed, all students' roles are extended to designers. The leader of the group in the design phase can be the same student as in the analysis phase, or the leader role can be assigned to another student at the beginning of the design phase. Also, in the same phase, the leader can be changed from time to time according to the supervisor requirements and the group decision.

CVAG differs from traditional hierarchical models, where the supervisor is the leader of the group and s/he has the role to manage and control the group at all the time. In CVAG, the supervisor focuses on mentoring the group  progress, and evaluating the work and the individual contributions. CVAG is designed to be used inside a programming class, therefore, the supervisor is typically the instructor for the class,  and  the person that prepares the programming task requirements. The group leader is one of the group members and, as we mentioned before, the group members can decide who is the leader and how the role can be changed from member to member  over the time. 

CVAG supports the supervisor with all features necessary to  facilitate his/her work to mentor and evaluate the work progress. S/He can trace the work progress and send  comments to the group.

\begin{table}[htbp]
\centering
\begin{tabular}{||l|p{6cm}||}
	\hline\hline
{\bf Role } & {\bf Definition}  \\
\hline
Author (driver) & Creates the actual program by performing the required transactions, such as: adding and modifying
classes and objects, creating methods, creating functions, defining variables/parameters, creating events, etc. \\
\hline
Leader & Assigns roles, changes roles, submits the deliverable to instructor, reviews the work progress, shares ideas about the members’ contribution, encourages members to finish on time.  \\
\hline 
Analyzer & Analyses the project requirements and create lists of objects, methods, functions, variables, events, etc.  \\
\hline
Designer & Designs a number of scenarios, the time line activities and events of the program, designs use-cases.  \\
\hline
Time keeper & Follows the time records for the work progress, notifies members of the required time for doing subtasks, notifies members of the deadline to finish the subtasks, and notifies members of the time to switch roles. \\
\hline
Explainer & Shares ideas and opinions. \\
\hline
Clarifier / paraphraser & Restates what other members have said to understand or clarify a message. \\
\hline 
Accuracy coach (Navigator) & Reviews the created program  by looking for errors, thinking about the overall structure of the code, finding necessary information and brainstorming with the group. \\
\hline
Integrator & Integrate members’ ideas and reasoning into a single position that everyone can agree on. \\
\hline\hline
\end{tabular}
\begin{flushleft}
\end{flushleft}
\caption{CVAG roles' definition}
\label{tab:role_def}
\end{table}

\subsection{Communication Features (Video, Audio and Chat)}
It is very important to ensure that  the group members maintain open communication at any point in time during the different phases of the project. Therefore, it is important to provide communication channels between users, especially in 
those situations where users have a virtual presence (e.g., an avatar) in the collaborative virtual environment. The different types of communication channels can be used in group meetings, orientations, or as an anytime  communication 
throughout the development of the project. Moreover, the group needs to communicate to share knowledge, make a conversation, make a decision, etc. Text chatting and audio\_video (video-conference) are two types of communication channels that are supported by CVAG.
\begin{itemize}
\item {\bf Chat:} is the basic communication channel that CVAG provides to keep the group contact. Chat, i.e., the ability to send immediate text messages between team members,  is a core feature in CVAG. Chat is a form of interaction that is immediate; a student can open a chat channel either to communicate with a specific other team member or with the entire team; the supervisor is also part of the communication process. CVAG allows the instructor to send a private message to a specific student in the group
or to broadcast information to the entire team. This type of communication channel can be useful to send information to others with details (e.g., send correct syntax of a programming statement, properties' values, the correct object name, etc.), transfer knowledge to others, transfer experiences or skills to others, etc.

\item {\bf Audio\_video (video-conference):} video conferencing is an ideal communication media to enable group meetings; it is encouraged media for both  small group and large group meetings. Students hang out in a video-conference, according to schedule times.  Video conferencing is often  not a good practice for quick communications during the development process---as it can be a cause of distraction~\cite{schmeil2009}. 
\end{itemize}

\subsection{Orientation and Group Meetings}
CVAG supports users (e.g., students) that are geographically dispersed; thus, it is critical for CVAG to provide opportunities for students to interact. We discussed in the previous sections the communication channels that CVAG provides to the users. These channels of communication are available at all  times. Nevertheless, efficiency could be gained by ensuring that communication is introduced in a well-structured context, and realized according to a specific time schedule. In CVAG, three types of meetings are designed for users (students and teachers), to facilitate their work and support them with the required 
 space to discuss, share, and transfer knowledge, experiences and skills. The three types of meetings differ in the time, purpose and/or communication channel, and they are discussed next.

\noindent 
{\bf Orientation:}
In the CVAG model, the group members are students who attend an introductory class to learn programming (e.g., an AP Computer Science Principles course). At the beginning of a programming course, students participate in 
an  \emph{orientation,}  prepared to allow students to:
\begin{itemize}
	\item Understand the philosophy and goals of CVAG,
	\item Understand the cooperative paradigm and learn about the requirements to learn programming in a collaborative setting,
	\item Become aware of teacher's expectations, and
	\item Understand the resources and features that CVAG provides to support their collaborative efforts.
\end{itemize}   
 The orientation is mandatory and realized using the audio-video communication channels---this is fundamental to allow remote students to connect and meet for the first time.

The orientation is a structured event. It starts by requiring each student to introduce her/himself, possibly following an established script. This is followed by a presentation from the instructor, describing 
 the project, goals, requirements and  his/her expectations. The orientation provides a space for students to present what they expect from taking the course and what their programming background is. 

\smallskip
\noindent 
{\bf Small Group Meetings:}
Small group meetings are expected to occur frequently throughout the development
of the project. We will distinguish two types of small group meetings. 
The first type includes \emph{regularly scheduled meetings}, typically occurring on a daily or 
bi-daily basis. A small group meeting can be held to discuss the work progress,  work that has just been completed, 
following tasks, members' roles,  problems that have been encountered and possible solutions. 
Small group meetings may involve the entire team or a subgroup (established by the instructor).
These meetings can make use of any communication channel---though video conferencing might be a preferred channel to
ensure a more direct and complete interaction. CVAG recommends these meetings to be scripted to ensure that all students
have the opportunity to communicate and that the goals of the meeting are properly met.

The second type of small group meeting includes impromptu meetings, typically including a small subset
of team members, and called on-the-fly to address an immediate need (e.g., a difficult bug). These meetings are typically
unscripted, brief, and conducted in close temporal proximity with the task prompting the need for communication. 
Text messages and text chat are preferred communication channels, as they are less distracting and enable communication
to occur while the students are still engaged in project activities (e.g., while they develop code)~\cite{schmeil2009}.

\smallskip
\noindent 
{\bf Large Group Meetings:}
Large group meetings in CVAG include all members of the team and they correspond to major 
milestones of the project. During a  large meeting, the group shares  announcements related 
to overall team achievements, such as published papers, awards, completion of major components, etc. The collaborative virtual model, in our case, will also include a large group meeting after each major learning milestone included in the project (e.g., submission of a part of the project, performance of an examination). Large group meetings are scheduled, scripted, and require the use of video-conferencing. 

\subsection{Project Management} 

\subsubsection{Resource Management}
In ARG, resource management means making suitable resources available to the right user. The available resources in ARG 
include items like lab time, office space, essential books, tutorials, etc. which are required to support the group's research project. 
Similarly, CVAG is designed to provide the proper resources to each student. Differently from ARG, CVAG customizes the resources provided based on the specific role(s) of the student.  For example, a student serving as a recorder  can open a writing board to collect the group's decisions during  the group work or in a meeting. Furthermore, CVAG's resources are designed to reflect the fact that the team operates in a virtual space.
The provided resources are automatically managed by the system  according to the user's role(s). 
Since the roles are dynamic,  the provided features will also change dynamically, reflecting  the changes in students' roles. A change in role  is not a trivial task---as it requires transferring complete control of features from one team member to another, making the use and appearance of the feature as seamless as possible.  
 For example, when a recorder writes on the virtual white board the group's decisions, the board is visible to the entire group (in read-only modality). When the role is assigned to another member, the new member should be given write authority to the virtual white board, and thus the ability to continue work on the same  decisions taken earlier by the group, as captured by the previous recorder.

\subsubsection{Consistency}
In a group of students that work together as members in a collaborative team, the issue of consistency arises and needs to be addressed. We refer to consistency as the guarantee that the created program seen by the team members are actually \emph{identical} and \emph{non-contradictory}. Identical views ensure that all team members are up-to-date with all changes being made (regardless of who made them). Lack of contradictions implies that all team members should be somehow prevented from making concurrent and conflicting changes to the same entity of the program. The consistency problem  arises from the fact that CVAG allows concurrent driver roles, thus enabling different students to actually create code belonging to the same project at the same time. 

In general, any implementation of CVAG should apply a specific algorithm to meet the consistency requirements. In section \ref{example}, we present \av\ as an example of an implementation of CVAG. The implemented algorithm uses a relatively straightforward mutual exclusion technique to prevent a conflict on specific parts being modified. Each Alice program contains a number of classes, objects, methods,  and functions. \av\ achieves consistency by placing mutual exclusion locks on each one of these entities whenever they are accessed by a student for creation/modification. Concurrent attempts to modify the same unit of code will cause an arbitrary sequencing of accesses; access counters can be used to ensure the lack of starvation. This approach reduces to the minimum the risk of inconsistencies (see also Section~\ref{example}).

\subsection{People and Resources Awareness}

Garcia et al. ~\cite{garcia2002move} define  \emph{awareness} in team work as an understanding of the activities of others, which provides a context of someone's own activities---making awareness core component of any collaborative environment. The distributed nature of the team in a CVAG setting makes the implementation of awareness even more critical and complex. 
In the design of CVAG, different types of features are provided  to support awareness. Not only CVAG promotes understanding of 
the membership of a team, but also supports the understanding of the roles and capabilities owned by each team member at any point in time. 
CVAG provides the team with a feature that allows each member to view the list of members, including the role of each member. In this case, each member will know what s/he can do and what are the other members' capabilities and limitations. Furthermore, CVAG supports \emph{Situational Awareness}, by allowing each student to understand their mutual ``position'' within the code, facilitating joint resolution of problems and reducing the risk of conflicts.

The group uses different types of communication channels to additionally create a level of
\emph{social awareness}. Students within the same group can communicate and interact throughout the project, facilitating the
establishment of social relationships. While other virtual environments rely on avatars to provide a form of users' embodiment,
CVAG enables situational awareness through text based notifications describing the specifics of each user's transaction.

\begin{figure*}[!htbp]
	\centering
	\includegraphics[width=1.2\textwidth]{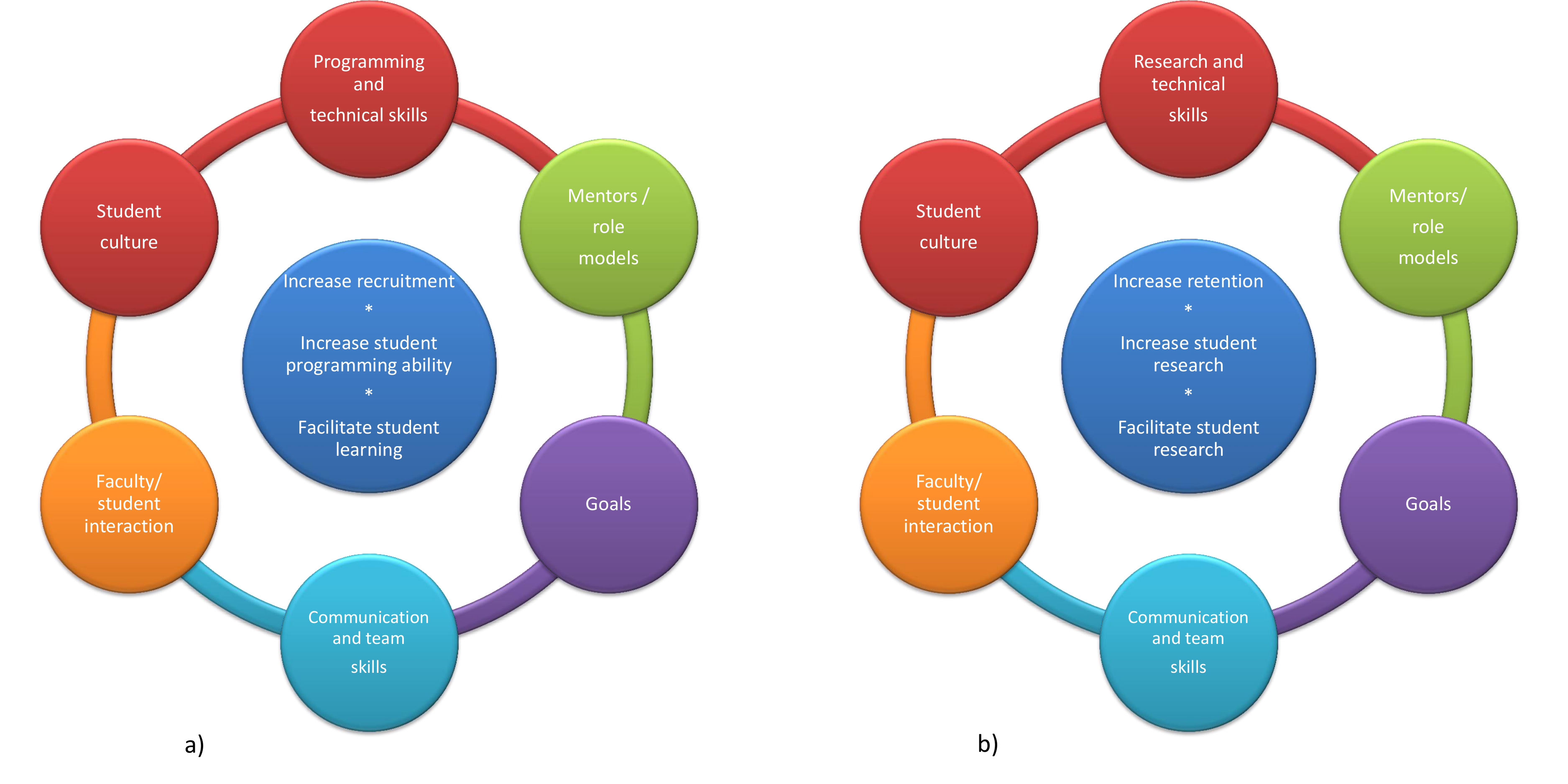}
	\caption{The CVAG (a) vs. ARG (b) components}
	\label{fig_CVAG_vs_ARG}
\end{figure*}

\section{CVAG vs. ARG}
The 
CVAG model has been designed as an extension and adaptation of the Affinity Research Group (ARG) model. ARG provides the opportunity to  students to learn, use, and integrate their knowledge and skills while  working on a joint 
 research project. In ARG, students work collaboratively and the model provides them with the structure
 necessary to gain knowledge and skills while achieving a collective project goal. The Collaborative Virtual Affinity Group model (CVAG) provides students with:
\begin{itemize}
\item The opportunity to learn programming in a collaborative virtual environment, working as part of a distributed team, 
\item The opportunity to gain the required experiences and skills to work in an affinity group,
\item A suitable environment for middle and high school students to learn programming,
\end{itemize}
In some ways, CVAG is, at the same time, both a generalization as well as a restriction of ARG. It serves as a restriction of ARG as it focuses exclusively on the creation and support of teams of students engaged in learning fundamental concepts of programming and computational thinking---while the original ARG model can be applied to a wide range of learning experiences.

On the other hand, CVAG provides an extension of ARG by relaxing the time and space restrictions of ARG.
Figure~\ref{fig_CVAG_vs_ARG} shows the main components of ARG and how they have been modified to meet the
goals of CVAG. The core goals of ARG, summarized in the middle circle in Figure~\ref{fig_CVAG_vs_ARG}b), are:
 increased retention, increased students research, and facilitated student research. The main goals of CVAG are, instead, shifted from research to learning (computer programming). 

In section~\ref{ARG_maped_comp}, we reviewed the different ARG components and discussed how they are mapped to the  CVAG model. The core purpose and the other five components of collaborative environments have been adapted for CVAG, in order  to make them suitable to be used in virtual environments. These types of changes require the model to be designed to handle geographically distributed teams of students. Moreover, CVAG defines a number of communication channels for students to facilitate interaction regardless of geographical location. On the contrary, these features are not necessary for ARG. All of the additional features which can be found in the design of CVAG  but not in the design of ARG are related to the virtual and distributed nature of the former. Therefore, consistency, persistence, low-level application communication, user communication, etc. are features that specifically belong to the realm of CVEs, and are thus absent in ARG. 

The other  important difference between ARG and CVAG is the target audience. The typical audience of
ARG is composed of graduate and undergraduate students engaged in research projects. The main target
audience of CVAG is composed of teachers and students at the middle and high school levels, engaged in homework and class projects while learning introductory programming.

 \section{A Concrete Implementation of CVAG}
\label{example}
The design of the CVAG model discussed in the previous sections has been validated in a concrete
implementation---built on the block-based interactive learning environment called \emph{Alice} \cite{alice}.

Alice was originally developed as an interactive programming environment, freely available, and designed to teach students the fundamental principles of programming---including relatively advanced features, like object oriented programming. Alice is extensively used at the  middle and high school levels. The design and implementation of new features and functionalities in Alice facilitate concurrent programming of 3D worlds by a \emph{distributed group of programmers,} following the principles of CVAG. We refer to the resulting environment as \emph{\av.} Thus, \av\ is an application to collaborate and to learn programming within  a collaborative virtual environment. By building this environment, we combine the benefits of using a well-established educational platform like Alice in teaching programming and the benefits from using collaborative models in programming and learning, e.g., increased technical  experiences and skills exchange, sharing of information, better program quality with fewer bugs, and enhanced student confidence. The first prototype of \av\ has been 
presented in~\cite{7044089} and focuses on generalizing Alice to support (virtual) pair programming.  
The second version of \av\ implements CVAG. The following is a summary of this second version of \av.

\subsection{The User Interface of \av} 
The graphical user interface (GUI) of Alice has been extended in \av. We explicitly emphasized the importance of making the novel features of \av\ as non-intrusive as possible, to reduce the learning curve of \av\ for students who are already familiar with Alice, and to avoid overwhelming inexperienced students with a complex interface. In particular, all of the code development tasks of Alice are available unchanged in \av, and they are presented to the users  with exactly the same visual interface---e.g., add an object to the world by selecting the \emph{``add object''} button and choosing the  object from gallery,  drag and drop statements into the methods, etc. 

The new components in the \av\ interface are related to the added functionalities of \av\ as a virtual collaborative environment. When the user starts the application, \av\ will prompt the user with a login dialog, requesting a user-name and a password. These two pieces of information have a dual role:
\begin{itemize}
	\item They provide security to the users (e.g., protecting the course-work that a student is developing from unintended accesses);
	\item The user name allows the identification of the user, necessary to place the user in the appropriate team and to 
		assign him/her the selected role(s).
\end{itemize}   
Information about users, teams and roles are maintained in an internal database within \av\ on the server side.

 \begin{figure}[t]
\centering
\includegraphics[width=.5\textwidth]{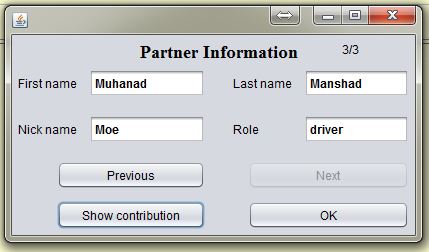}
\caption{The group member information. The figure show that the group has three members (as the number at the top right side of the window) and the user show the information of the third member}
\label{partnerInfo}
\end{figure}

\begin{figure}[t]
\centering
\includegraphics[width=.65\textwidth]{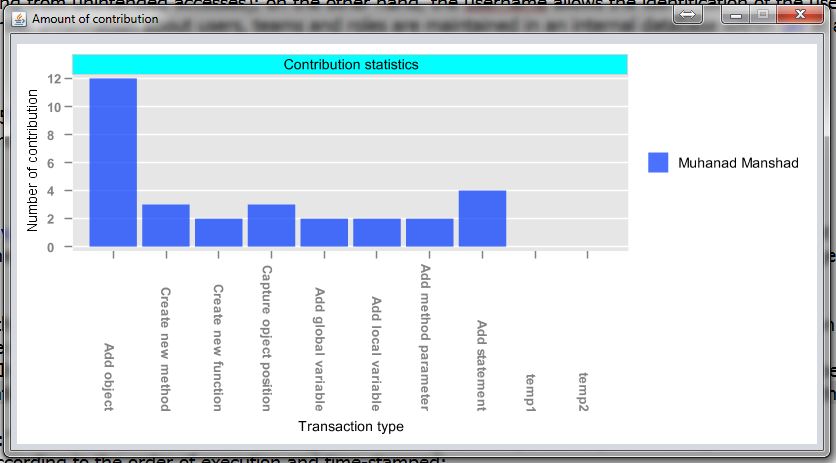}
\caption{The contribution of a user in the team}
\label{userCont}
\end{figure}

\begin{figure}[t]
\centering
\includegraphics[width=.65\textwidth]{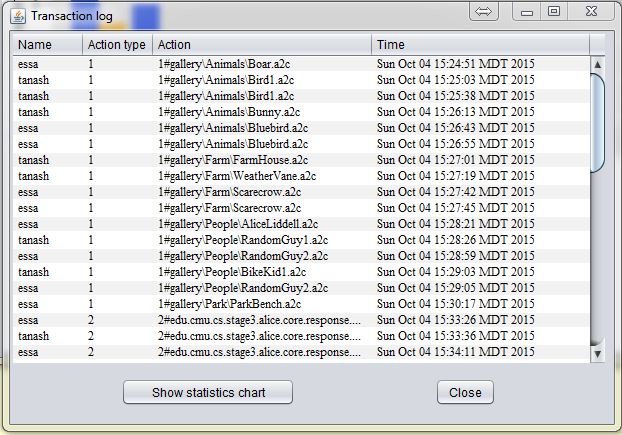}
\caption{The transaction log show all transaction for all user}
\label{transLog}
\end{figure}

In the main window of \av, the user is presented with the same interface as Alice, with the three new components. The first component is a new menu item in the menu bar, called \emph{Collaboration}. The second component is a new button labeled \emph{Chat}. The third component is a text message field located in the tool bar. 

The Collaboration item is a menu with  four options:
 \begin{enumerate}
 \item \emph{``Chat:''} this option establishes a chat window with the other team members and the user in the window can establish a video conference (synchronized audio and video) with other team members; the student can also request the chat using the shortcut button on the tool bar;
  \item \emph{``Partner Information:''} this option allows the user to retrieve basic information about the other team members (Figure~\ref{partnerInfo}),
  	including his/her identity, his/her current role(s), and a list of the most recently performed operations by the  partner. The ``Show contribution'' button displays a chart of the  transactions performed by the partner, organized by type of
  	transaction  (Figure~\ref{userCont});

  \item \emph{``Transaction Log:''} this option provides access to a log file containing all the operations performed by all team members, sorted according to the order of execution and time-stamped (Figure~\ref{transLog}). The user  can also request a chart containing the statics about all the transactions performed by the entire team (Figure~\ref{fig_stat}); 
  
  \item \emph{``User Profile:''} this option allows the user to review his/her own profile and manage the information in it (e.g., modify password); the user profile allows the user to also change his/her role, if the supervisor has granted permission to do so.
 \end{enumerate}

\subsection {\it Users Information}  \av\ keeps the required information in an XML database. The database has the basic information about each programmer: user name, password, roles, etc. and the basic information about the project that each team is currently working on. Users information is saved on the server side. When the user enters his/her information to login (user name and password), the \av\ client sends a message to the server; if the server finds the user, it will return an acknowledgment containing the complete user record (e.g., the user role), otherwise, the user receives an error message. Please note that the users are created by the instructor, in order to provide a control on the format of created teams.

\begin{figure*}[htbp]
\centering
\includegraphics[width=.75\textwidth]{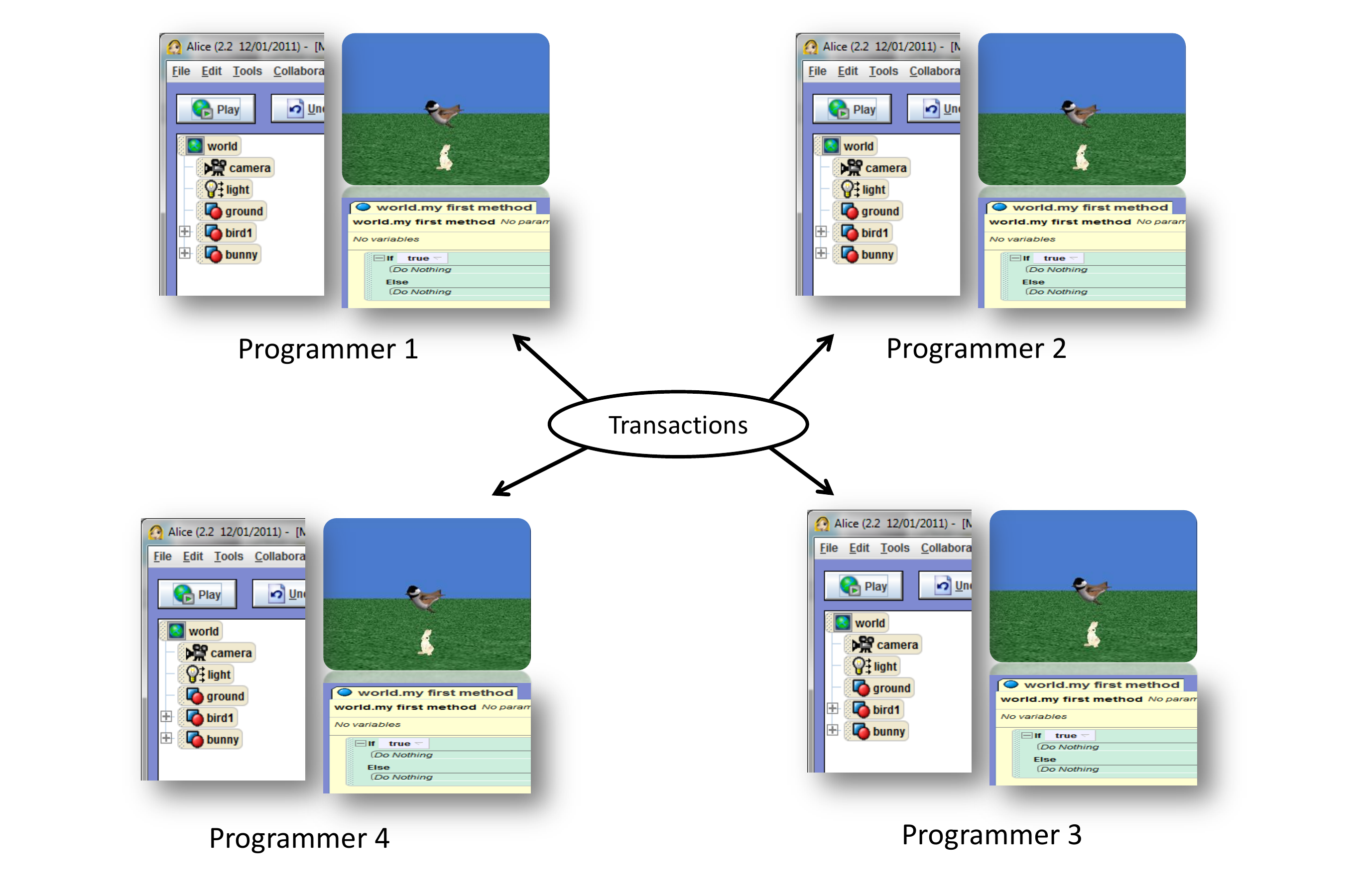}
\caption{The transaction broadcast to all members}
\label{transBroadcast}
\end{figure*}

\subsection{Application-level Communication Mechanisms} 
\av\ is a collaborative virtual environment that provides synchronous view of a shared 3-D world on  different workstations (users' sides), connected through the Internet. In order to ensure synchronous behavior as well as to provide all the communication channels described  earlier (e.g., notifications of changes, audio/video communications), \av\ needs a communication layer that supports the exchange of events among the remote partners. The transfer of events between users is realized through \emph{RabbitMQ}~\cite{rabbitmq}. RabbitMQ is an open source message broker software, that provides a reliable method to send and receive messages. The Advanced Message Queuing Protocol (AMQP) implemented in RabbitMQ is a suitable protocol for \av. Two team members' sessions are connected by an exchange message queue; each team member will broadcast a message to other team members. Figure ~\ref{transBroadcast} show four members in a group and how they are connected. The interface for the four members are identical and all of them see the same components---i.e., world scene, project browser, methods, etc.

\begin{figure*}[t]
\centering
\begin{subfigure}[t]{.45\textwidth}
\includegraphics[width=\textwidth]{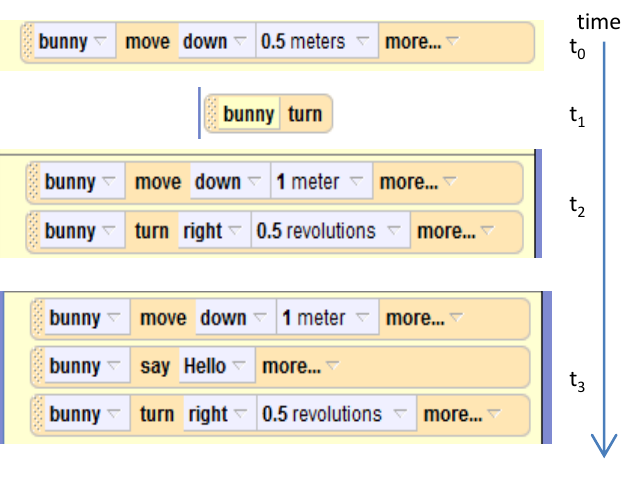}
\caption{Driver I Transaction}
\label{Div1_trans}
\end{subfigure}
\begin{subfigure}[t]{.45\textwidth}
\includegraphics[width=\textwidth]{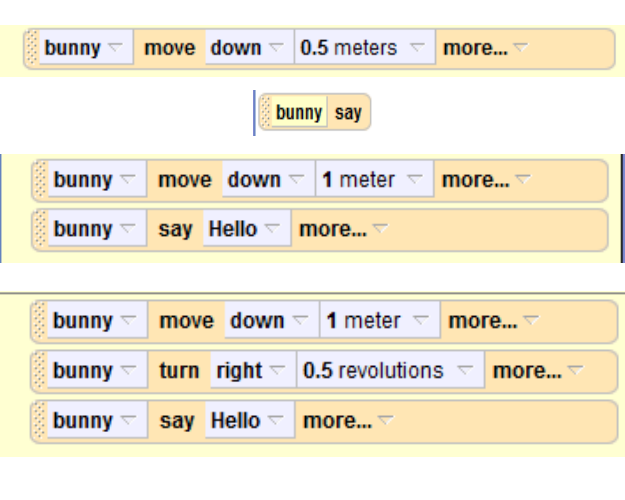}
\caption{Driver II Transaction}
\label{Div2_trans}
\end{subfigure}
\caption{The transaction executed at two drivers sides according to the time. At time $t_{0}$, the two drivers have the same state. At time $t_{1}$, both driver drag a statement to add it to the method, after that the statement added as in time $t_{2}$, then a transaction message sent to add the statement on the partner side, after receive the message which say to insert the statement at index 1 (after the first statement), the final state on both side as in time $t_{3}$}
\label{fig_cons}
\end{figure*}

\subsection{Consistency Implementation} Let us further elaborate on the two aspects of consistency. Consistency can be lost as result of two potential situations---resulting in  two or more team members in the group seeing different versions of the Alice 3D world:
\begin{enumerate}
	\item The first situation occurs  when a transaction performed by one team member is not received by other team members (and, thus, not reflected in the 3D worlds seen  by some members of the team).
	\item  The second situation derives from  conflicting changes. Two or more team members  (e.g., all acting as drivers) may apply conflicting modifications to the same object of the world; for example, two students may   both drag and drop a distinct statement in the same method. The  order of statements in the method is important, and there is a danger that such order will differ in the two 3D worlds. For example, Figure~\ref{Div1_trans} and Figure~\ref{Div2_trans} show the transactions executed by both drivers on each side, resulting in inconsistent methods. Similar problems may occur when two students modify the same property of an entity at the same time---depending on the speed of communication, the 3D worlds may result in different values of such property (or, in a less ``threatening'' case, one of the changes may be lost).
\end{enumerate} 

The first situation may occur for various reasons,  ranging from shutdown of one of the machines to transient communication problems. The solution for this problem relies on the use of  acknowledgment messages. Several well-established schemes have been proposed in the distributed computing literature to address unreliable communication, through the use of delays and acknowledgment messages. The current implementation of \av\ makes use of a relatively simple approach: after each transaction is sent to all machines,  an acknowledgment message should be sent back. If no acknowledgment message is received within a given time window, a flag will be set in the log file to show that the transaction has not been properly executed and  all team members are alerted of  the problem. In this case, the student may have to wait until the remote 3D world is properly updated. 

The second situation can be resolved using lock messages, generated by the drivers before performing a transaction and associated to the entity which is the subject of the transaction (e.g., driver 1 may generate a lock message associated to method1 before performing a drag and drop of a statement in method1). The actual transaction will be performed only after locking the local entity and receiving an acknowledgment message indicating that the entity has been successfully locked on all remote machines. The locking may fail, e.g., if the entity is already locked by the other team member; in such a case, the current entity will be released, the transaction will not occur and the driver will be notified. If the lock is successful, then the transaction will be executed on the current entity and a message sent to the remote machines to request the proper update of the entity. The receipt of the transaction will also correspond to the unlocking of the remote entity. If drivers fail to complete a transaction because they noticed that two or more of the other students are working on the same entity, then they can communicate to coordinate the work.  

\section{Conclusion and Future Work}
In this paper, we presented a generalization and adaptation of the Affinity Research Group (ARG) model, which enables its application and use to support learning introductory programming in a collaborative virtual environment. The paper discussed the different components of the resulting model, called CVAG, comparing and contrasting it with ARG and with other fundamental principles underlying the design of collaborative virtual environments. The resulting model has been implemented in a concrete learning environment, \av.

The current research directions are aimed at completing the development of a robust version of \av\ that can be disseminated to the public; we are also exploring the formal assessment of \av\ in both a classroom setting (e.g., in a course adopting the College Board Computer Science Principles curriculum) as well as in an after-school program for middle school students.
We are also exploring how to lift the principles of CVAG to make them applicable to other introductory programming learning environments, such as Scratch \cite{scratch} and AppInventor \cite{appinventor}.



\bibliographystyle{IEEEtran}

%


\end{document}